\documentclass[prc,twocolumn,nofootinbib,superscriptaddress,showpacs,aps]{revtex4-1}

\usepackage{amsmath}
\usepackage{amssymb}
\usepackage{graphicx}
\usepackage{dcolumn}
\usepackage{bm,braket}
\usepackage{hyperref}
\usepackage{isotope}
\usepackage{color}
\usepackage{multirow}
\usepackage[activate={true,nocompatibility},final,tracking=true,
kerning=true,spacing=true,factor=1100,stretch=10,shrink=10]{microtype}

\usepackage[all]{nowidow}
\hypersetup{
     colorlinks   = true,
     linkcolor = cyan,
     citecolor = cyan,
     menucolor = black,
     urlcolor = cyan  
}

\begin{document}
\renewcommand{\vec}[1]{\mathbf{#1}}
\newcommand{\clebschG}[6]{\mathcal{C}_{#1 #2 #3 #4}^{#5 #6}}  
\newcommand{\kf}{k_{\rm F}}
\newcommand{\clebschGA}[6]{
  \left(
  \begin{array}{cc|c}
  #1 & #3 & #5 \\
  #2 & #4 & #6
  \end{array}
  \right)
}
\newcommand{\sixJsymbsixA}[6]{
	\begin{Bmatrix}
  #1 & #3 & #5 \\
  #2 & #4 & #6
	\end{Bmatrix}
}  
\newcommand{\nineJsymbA}[9]{
	\begin{Bmatrix}
  #1 & #4 & #7 \\
  #2 & #5 & #8\\
  #3 & #6 & #9 
	\end{Bmatrix}
}

\title{Neutron matter from chiral two- and three-nucleon calculations up to N$^3$LO}

\author{C.\ Drischler}
\email[Email:~]{christian.drischler@physik.tu-darmstadt.de}
\affiliation{Institut f\"ur Kernphysik, Technische Universit\"at Darmstadt, 64289 Darmstadt, Germany}
\affiliation{ExtreMe Matter Institute EMMI, GSI Helmholtzzentrum f\"ur Schwerionenforschung GmbH, 64291 Darmstadt, Germany}

\author{A.\ Carbone}
\email[Email:~]{arianna@theorie.ikp.physik.tu-darmstadt.de}
\affiliation{Institut f\"ur Kernphysik, Technische Universit\"at Darmstadt, 64289 Darmstadt, Germany}
\affiliation{ExtreMe Matter Institute EMMI, GSI Helmholtzzentrum f\"ur Schwerionenforschung GmbH, 64291 Darmstadt, Germany}
 
\author{K.\ Hebeler}
\email[Email:~]{kai.hebeler@physik.tu-darmstadt.de}
\affiliation{Institut f\"ur Kernphysik, Technische Universit\"at Darmstadt, 64289 Darmstadt, Germany}
\affiliation{ExtreMe Matter Institute EMMI, GSI Helmholtzzentrum f\"ur Schwerionenforschung GmbH, 64291 Darmstadt, Germany}
 
\author{A.\ Schwenk}
\email[Email:~]{schwenk@physik.tu-darmstadt.de}
\affiliation{Institut f\"ur Kernphysik, Technische Universit\"at Darmstadt, 64289 Darmstadt, Germany}
\affiliation{ExtreMe Matter Institute EMMI, GSI Helmholtzzentrum f\"ur Schwerionenforschung GmbH, 64291 Darmstadt, Germany}
\affiliation{Max-Planck-Institut f\"ur Kernphysik, Saupfercheckweg 1, 69117 Heidelberg, Germany}

\begin{abstract} 

Neutron matter is an ideal laboratory for nuclear interactions derived
from chiral effective field theory since all contributions are
predicted up to next-to-next-to-next-to-leading order (N$^3$LO) in the
chiral expansion. By making use of recent advances in the partial-wave
decomposition of three- nucleon (3N) forces, we include for the first
time N$^3$LO 3N interactions in many-body perturbation theory (MBPT)
up to third order and in self-consistent Green's function theory
(SCGF). Using these two complementary many-body frameworks we provide
improved predictions for the equation of state of neutron matter at
zero temperature and also analyze systematically the many-body
convergence for different chiral EFT interactions. Furthermore, we
present an extension of the normal-ordering framework to finite
temperatures.  These developments open the way to improved
calculations of neutron-rich matter including estimates of theoretical
uncertainties for astrophysical applications.

\end{abstract}


\maketitle

\section{\label{sec:intro}Introduction}

Progress in chiral effective field theory (EFT) for nuclear
forces~\cite{Epel09RMP,Mach11PR} and advances in many-body
theory~\cite{Soma14GGF2N3N,Bind14CCheavy,Holt14Ca,Lahd13LEFT,Hage14rev,Sign14BogCC,Dikm15NCSMSM,Herg16IMSRG}
offers new paths to systematically improvable calculations of nuclear
many-body systems~\cite{Hamm12RMP,Hebe15ARNPS}. In recent years
infinite nuclear matter has been studied based on chiral EFT
interactions within various frameworks like many-body perturbation
theory (MBPT)~\cite{Hebe11fits,Cora14nmat,Well14nmtherm,Dris16asym},
in-medium chiral perturbation theory~\cite{Holt13PPNP},
self-consistent Green's function (SCGF) framework~\cite{Carb13nm},
coupled-cluster theory~\cite{Hage14ccnm}, the Brueckner-Hartree-Fock
approach~\cite{Kohn13gmatchiral,Isau16pnmdin} and Quantum Monte
Carlo methods~\cite{Geze13QMCchi,Wlaz14QMC,Rogg14QMC,Lynn15QMC3nf}.

So far, the employed chiral EFT nucleon-nucleon (NN) and three-nucleon
(3N) interactions in these calculations were all derived within
Weinberg's power counting scheme~\cite{Epel09RMP,Mach11PR}. Here the
leading 3N forces appear at next-to-next-to-leading order (N$^2$LO)
and contain two unknown low-energy couplings, $c_D$ and $c_E$, which
need to be determined by fits to few- or many-body observables. In
contrast, subleading 3N forces at N$^3$LO do not contain any new
low-energy couplings~\cite{Bern083Nlong,Bern113Nshort} and are thus
completely predicted. Hence, including these contributions in
calculations offers the possibility to probe systematically the
validity of chiral power counting in nuclear systems and to provide
estimates of theoretical uncertainties.

Full N$^3$LO calculations of neutron matter were first performed in
Refs.~\cite{Tews13N3LO,Krue13N3LOlong}. These works showed that 3N
forces at N$^3$LO provide surprisingly large contributions to the
equation of state especially in symmetric matter. Similar results were
found for few-body systems in Ref.~\cite{Hebe15N3LOpw}. These findings
raise fundamental questions concerning the convergence of the chiral
expansion for 3N forces within the employed regularization and power
counting scheme.

Generally, the treatment and inclusion of 3N forces is still a
challenge in many-body calculations. In particular, due to the
complexity and rich analytical structure of 3N forces at
N$^3$LO~\cite{Bern083Nlong,Bern113Nshort} so far it was possible to
include effects from 3N interactions only at the Hartree-Fock level in
Refs.~\cite{Tews13N3LO,Krue13N3LOlong}. While this approximation is
expected to be reliable for neutron matter, higher-order terms in the
many-body expansion are expected to become significant as soon as the
proton fraction becomes sufficiently large. In the present paper we
address this issue by making use of two recent advances: a) the
development of a novel framework that makes it possible to compute
matrix elements of 3N interactions in a partial-wave momentum
basis~\cite{Hebe15N3LOpw} and the availability of matrix elements up
to N$^3$LO and large model spaces, and b) the development of a novel
normal-ordering framework based on partial-wave matrix
elements~\cite{Dris16asym} that allows to systematically include these
3N interactions in calculations of nuclear matter for arbitray isospin
asymmetry.  By combining these two advances it is now possible to
include general 3N forces that are available in form of plane-wave
partial-wave matrix elements and to treat 3N forces on the same
footing as NN forces in the many-body expansion. Furthermore, these
developments play an important role in view of future calculations
that will employ simultaneous evolution of NN and 3N interactions in a
momentum basis via similarity renormalization group
techniques~\cite{Hebe12msSRG,Hebe2013nmsrg}.

In this paper we will exploit and combine these new capabilities and
perform improved calculations of neutron matter up to N$^3$LO in MBPT
and SCGF. We benchmark results of these two complementary many-body
framework against each other and present a generalization of the
normal-ordering framework to finite temperatures. The extension of the
present N$^3$LO calculations to arbitrary proton fractions is in principle
straightforward but requires reliable fit values for the low-energy
couplings $c_D$ and $c_E$ at this order~\cite{Skibi113HN3LO3N, Gola14n3lo}. 
This is work in progress.  In neutron
matter these short-range and mid-range topologies do not contribute
within the employed regularization scheme.

The paper is organized as follows. In Sec.~\ref{sec:calc_details} we
specify the set of employed chiral EFT Hamiltonians and describe the
novel normal-ordering framework that allows to include general 3N
interactions in calculations of nuclear matter. In addition we briefly
discuss the many-body frameworks we used for our calculations. In
Sec.~\ref{sec:results} we present our results based on three different
sets of Hamiltonians, with a special focus on the effects of 3N forces
beyond the Hartree-Fock approximation.  Furthermore, we analyze the
many-body convergence in MBPT by comparing with SCGF results. Finally
we present a generalization of our normal-ordering framework to finite
temperatures and benchmark results for the energy against exact
Hartree-Fock results. In Sec.~\ref{sec:outlook} we conclude with a
summary and an outlook.

\section{\label{sec:calc_details}Calculational details}

\subsection{\label{subsec:chiral_ham}Chiral EFT Hamiltonians}

We consider unevolved NN and 3N forces up to N$^3$LO and calculate the  energy
per particle of infinite neutron matter in the frameworks of MBPT and SCGF.
The Hamiltonian takes the form
\begin{equation}
\label{eq:ham}
H = T + V_\text{NN} + V_\text{3N} + \ldots \,,
\end{equation}
where $T$, $V_\text{NN}$ and $V_\text{3N}$ denote the kinetic energy,
the NN and 3N intercations, respectively. So far, in most calculations
of nuclear matter NN and 3N forces were not included consistently up
to the same order in the chiral expansion due to the complex structure
of 3N forces at N$^3$LO~\cite{Bern083Nlong,Bern113Nshort}. Only
recently an efficient partial-wave decomposition of these
contributions was developed in Ref.~\cite{Hebe15N3LOpw}. In
Refs.~\citep{Tews13N3LO,Krue13N3LOlong} the N$^3$LO 3N contributions
were evaluated exactly for neutron matter and symmetric nuclear matter
in Hartree-Fock approximation. It was somewhat unexpected that the
subleading 3N forces provide significant contributions to the
energy. The findings suggest that it is mandatory to investigate these
contributions more systematically by including higher-order effects in
the many-body expansion.

We note that, considering only NN and 3N forces at N$^3$LO in
Eq.~(\ref{eq:ham}) is still not fully consistent in the chiral
expansion. In fact, four-nucleon (4N) forces also contribute at this
order. However, Ref.~\citep{Kais124N,Tews13N3LO,Krue13N3LOlong}
demonstrated that the 4N contributions to the energy in neutron matter
in the Hartree-Fock approximation are very small compared to the
overall uncertainty, \mbox{$E_\text{4N}/N \sim -180$~keV} at
saturation density.  Therefore, 4N contributions only lead to a small
shift for all Hamiltonians and do not affect the relative comparison
of MBPT and SCGF. Consequently, if not stated otherwise, we neglect 4N
(and higher-body) contributions in Hamiltonian~\eqref{eq:ham} and
focus on the improvement of subleading 3N forces.

Normal-ordering with respect to a reference state is a well-known
method to include 3N contributions in terms of density-dependent
effective NN forces, which can then be directly included in NN
frameworks. Usually, the remaining residual 3N Hamiltonian leads to
small contributions in pure neutron matter and is thus neglected (see,
e.g., Ref.~\cite{Hage14ccnm}). Following
Refs.~\cite{Hebe10nmatt,Holt10ddnn} we obtain the effective NN
interaction $\overline{V}_\text{3N}^\text{as}$ by summing one particle
over the occupied states of the reference state, i.e.,
\begin{equation}
\label{eq:Veff_formal}
\overline{V}_\text{3N}^\text{as} = \text{Tr}_{\sigma_3} \int \frac{d\vec{k}_3}{(2\pi)^3} \mathcal{A}_{123} V_\text{3N} \, n_{\vec{k}_3} \bigg|_{\text{nnn}} \, ,
\end{equation}
with the momentum-distribution function $n_\vec{k}$ and
$\mathcal{A}_{123}$ is the antisymmetrizer. At zero temperature it is
common to approximate the distribution function by the free Fermi gas
function $n_\vec{k} = \Theta \left( k_{\rm F} - |\vec{k}| \right)$,
with Fermi momentum $k_{\rm F}$. It was demonstrated that the
inclusion of correlations in the reference state leads to small
effects in observables~\cite{Carb14SCGFdd}. In this article, we also
discuss the extension of the normal-ordering framework to finite
temperatures.

The 3N interactions $V_\text{3N}$ are regularized using non-local
regulators of the form $f_\text{R}(p,q)=\exp [-((p^2+ 3 q^2/4)/
\Lambda_{\text{3N}}^2)^{4}]$ with respect to the Jacobi momenta
$p,q$. In the literature, Eq.~\eqref{eq:Veff_formal} has been first
evaluated directly based on the operatorial form of the 3N forces at
N$^2$LO~\cite{Holt10ddnn,Hebe10nmatt, Hebe11fits}. Since this
procedure becomes rather involved for subleading 3N forces, so far
only leading 3N interactions could be considered in this approach. One
way to solve this is to make use of the recently developed
partial-wave decomposition of the 3N interactions~\cite{Hebe15N3LOpw}
and evaluate Eq.~\eqref{eq:Veff_formal} in a partial-wave momentum 
basis of the form
\begin{equation}
\Ket{pq\alpha} \equiv \Ket{pq; \left[(LS)J \left(l\frac{1}{2}\right)j \right] \mathcal{J} \left(T\frac{1}{2}\right)\mathcal{T}}\, .
\end{equation}
The quantum numbers $L$, $S$, $J$, and $T=1$ (for neutron matter)
denote the relative orbital angular momentum, spin, total angular
momentum, and isospin of particles $1$ and $2$ with relative momentum
$p$. The quantum numbers $l$ and $j$, respectively, are the orbital
angular momentum and total angular momentum of particle $3$ relative
to the center of mass of the pair with relative momentum $p$. The
quantum numbers $\mathcal{J}$ and $\mathcal{T}=3/2$ define the total
3N angular momentum and isospin. The 3N matrix elements are provided
by Ref.~\citep{Hebe15N3LOpw} with total three- and two-body quantum
numbers $\mathcal{J} \leq 9/2$ and $J \leq 6$, respectively. The size
of this model space is sufficient to ensure convergence for
calculations of nuclear matter in the Hartree-Fock
approximation~\cite{Hebe15N3LOpw,Dris16asym} (see also
Sec.~\ref{subsec:NO_finT}). The resulting effective NN interaction is
then added to the NN interactions:
\begin{equation}
V_\text{NN+3N}^{\text{as}} = V_\text{NN}^{\text{as}} + \zeta \overline{V}_\text{3N}^{\text{as}} \, .
\end{equation}
We refer to Ref.~\citep{Hebe10nmatt,Carb14SCGFdd,Dris16asym} for
detailed discussions on the combinatorial normal-ordering factor
$\zeta$. We also note that the summation in Eq.~\eqref{eq:Veff_formal}
results in a dependence of $\overline{V}_\text{3N}^\text{as}$ on the
total momentum $\vec{P}$ of the two particles, which is not the case
for free-space NN forces due to Galilean invariance. This additional
momentum makes the effective NN potential \eqref{eq:Veff_formal}
computationally involved. Commonly, the approximation $\vec{P}=0$ is
applied, e.g., in Ref.~\cite{Holt10ddnn,Hebe10nmatt,Carb13nm}.  In
Ref.~\cite{Dris16asym}, an additional approximation that averages over
all directions of $\vec{P}$ opposed to $\vec{P}=0$ is studied. It is
shown that the resulting 3N Hartree-Fock energies are in reasonable
agreement in particular below saturation density. Since the dependence
on $\vec{P}$ is currently not implemented in the SCGF code and since
we focus on the benchmark of MBPT to this nonperturbative method we
focus here on the $\vec{P}=0$ approximation for
$\overline{V}_\text{3N}^{\text{as}}$. Finally, we note that once
reasonable fit values for $c_D$ and $c_E$ are available at N$^3$LO,
the described methods can be directly applied beyond neutron matter.

\subsection{\label{subsec:many_body_app}Many-body frameworks}

\begin{figure*}[t]
\includegraphics[page=1,scale=0.95,clip]{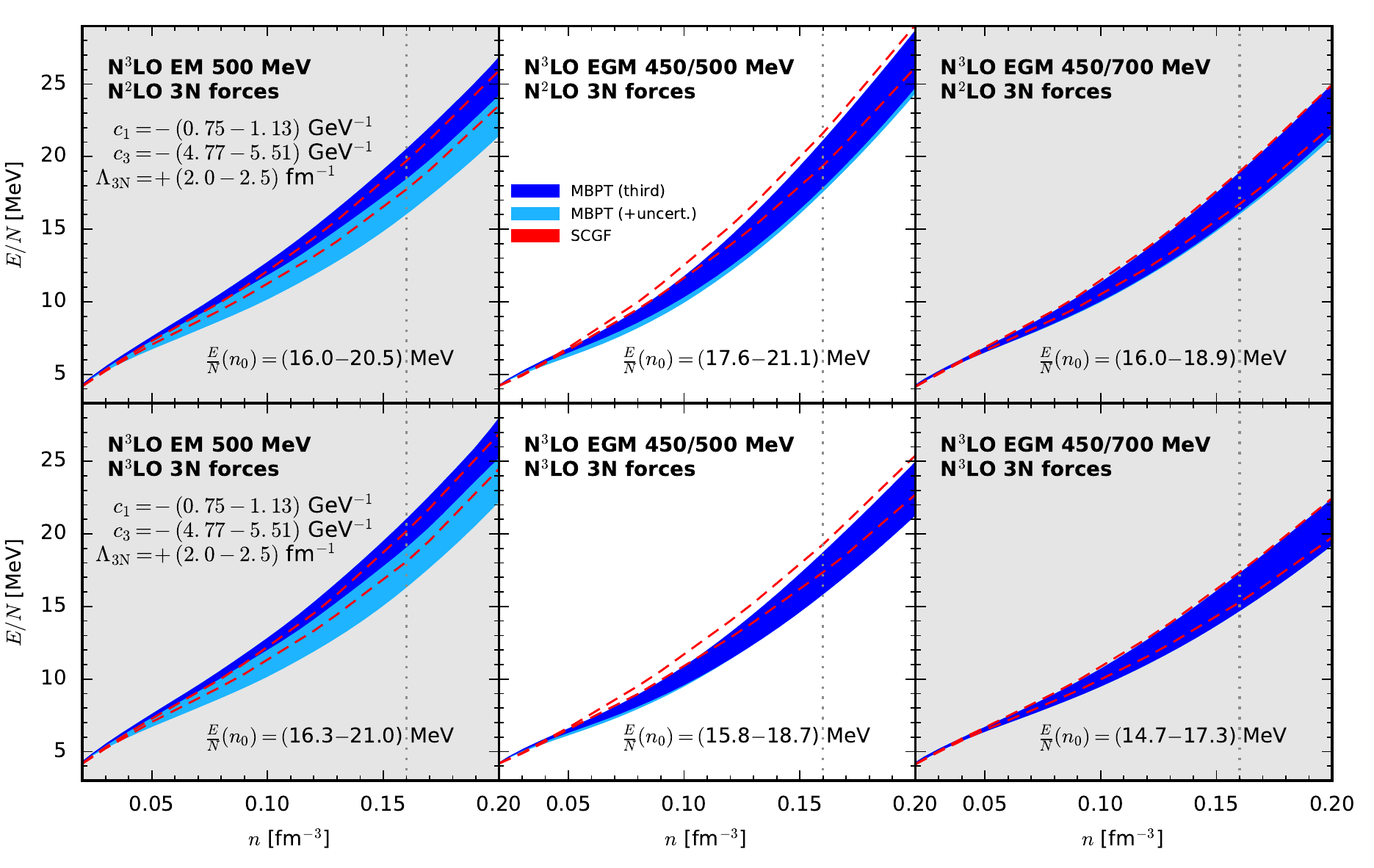}
\caption{\label{fig:EOS}(Color online)
The energy per particle in neutron matter for three different N$^3$LO 
NN potentials with N$^2$LO (top) and N$^3$LO (bottom) 3N forces,
respectively. The uncertainty bands are due to the given $c_i$ and 3N
cutoff variation.  For the MBPT results, we consider in addition the
maximum range of third-order calculations with a free and a
Hartree-Fock spectrum (dark-blue band) plus the change from a
second-order calculation with a Hartree-Fock spectrum, which is
indicated by the light-blue extension of the pure third-order
uncertainty band. The two bands together define the total uncertainty
estimate of MBPT. The region between the two red-dashed lines denotes
the uncertainty band of the SCGF method, which we do not fill for a
better view. In each panel the energy range at saturation density
obtained in MBPT is given.}
\end{figure*}

We calculate the energy per neutron at zero temperature up to third
order in MBPT. The following notation is used to distinguish
interaction energies and total energies at a given order in
perturbation theory:
\begin{subequations}
\begin{align}
\frac{E^{ \text{(HF)} }}{N} &= \frac{T}{N} +\frac{E^{(1)}}{N} \, ,\\
\frac{E_{\text{tot}}^{(2)}}{N} &= \frac{E^{(\text{HF})}}{N} +\frac{E^{(2)}}{N}\, ,\\
\frac{E_{\text{tot}}^{(3)}}{N} &= \frac{E_{\text{tot}}^{(2)}}{N} +\frac{E^{(3)}}{N}  \, .
\end{align}
\end{subequations}
Particle-hole contributions are neglected at third-order similarly to
Refs.~\citep{Hebe10PRL,Krue13N3LOlong,Dris16asym}. In order to
estimate the uncertainties due to neglected higher-order contributions
we perform calculations with a free and a Hartree-Fock single-particle
spectrum. We refer to Ref.~\citep{Hebe10nmatt,Dris16asym} for details
of the calculation.  We assess the many-body convergence
order-by-order by comparing to SCGF. In the SCGF method, the energy
per neutron is calculated nonperturbatively via knowledge of a dressed
one-body Green's function~\cite{Dick04PPNP}. The energy is obtained in
the so-called ladder approximation, where an infinite sum of
particle-particle and hole-hole diagrams is
performed~\cite{Rios08hotscgf,Soma093ntherm}. Similar to the MBPT
calculations, particle-hole contributions are neglected. The SCGF
approach has been recently extended to self-consistently include 3N
forces~\cite{Carb13SCGF3B}. In this extension, the ladder resummation
and the self-energy are redefined incorporating normal-ordered 3N
terms with respect to a dressed reference state. Residual 3N
contributions are also neglected in this approach. In this extended
approach, the modified sum rule to obtain the total energy per
particle in neutron matter reads~\cite{Carb13SCGF3B}:
\begin{equation}
\label{eq:gmk}
\frac{E}{N}= \frac{2}{n}\int\frac{{\rm d}{\bf k}}{(2\pi)^3}\int\frac{{\rm d}\omega}{2\pi}\frac{1}{2}\left\{\frac{k^2}{2m}+\omega\right\}A(k,\omega)f(\omega) - \frac{\langle W\rangle}{2} \,,
\end{equation}
where $n$ the total density of the system and $f(\omega)$ corresponds
to the Fermi-Dirac distribution function.  $A(k,\omega)$ is the
spectral function; this quantity gives the probability of adding or
removing a particle with momentum $k$ which causes an excitation in
energy ${\rm d}\omega$ in the many-body system. $\langle W\rangle$ is
the expectation value of the 3N operator (see Ref.~\cite{Carb14SCGFdd}
for details).  Throughout the paper we will refer to
Eq.~\eqref{eq:gmk} as $E_{\rm SCGF}/N$. The present implementation of SCGF is not
capable of treating the appearance of pairing below a critical
temperature, for this reason calculations are always performed at
finite $T$. The pairing instability
does not affect the MBPT calculations because the energy diagrams are
evaluated directly, for which the pairing singularity is
integrable. The zero-temperature results in SCGF are extrapolated
using the Sommerfeld expansion~\cite{Rios08hotscgf}. In this expansion, the
energy can be written as a quadratic expansion in terms of
$T/\varepsilon_{\rm F}$, where $\varepsilon_{\rm F}$ is the Fermi
energy, as long as $T/\varepsilon_{\rm F}\ll 1$. A more sophisticated
computational method to numerically extrapolate self-energies,
spectral functions and thermodynamical properties from finite to zero
temperature has been recently presented in
Ref.~\cite{Ding16scgfpair}.

In order to extend the effective NN interaction
$\overline{V}_\text{3N}^\text{as}$ to finite temperatures, we extend
the framework presented in Ref.~\cite{Dris16asym} and evaluate
Eq.~\eqref{eq:Veff_formal} at finite temperature using the general
Fermi-Dirac distribution function, $n_\vec{k} = \left[ \exp(
\beta(\varepsilon_\vec{k}-\mu) ) + 1 \right]^{-1}$.  Given a total
density $n$, we compute the chemical potential $\mu(n)$ by solving the
non-linear density relation
\begin{equation}
n = \frac{1}{\pi^2} \int \limits_0^\infty dk \, k^2 n_\vec{k}(\mu) \, .
\end{equation}
We consider here the free single-particle energy, i.e.,
\mbox{$\varepsilon_\vec{k} = \vec{k}^2/(2m)$}. Higher-order
corrections to the self energy include contributions from the
effective NN potential itself and would require thus an involved
self-consistent solution for the spectrum. It has been shown in
Ref.~\cite{Carb14SCGFdd} that the energy per particle in pure neutron
matter shows only at higher densities a dependence on the momentum
distribution used in Eq.~\eqref{eq:Veff_formal}. Such high densities
are not considered in this work, but it will be important to check
this approximation at high temperatures.

\section{\label{sec:results}Results}

\subsection{Comparison of MBPT and SCGF}

\begin{figure*}[t]
\includegraphics[page=1,scale=0.95,clip]{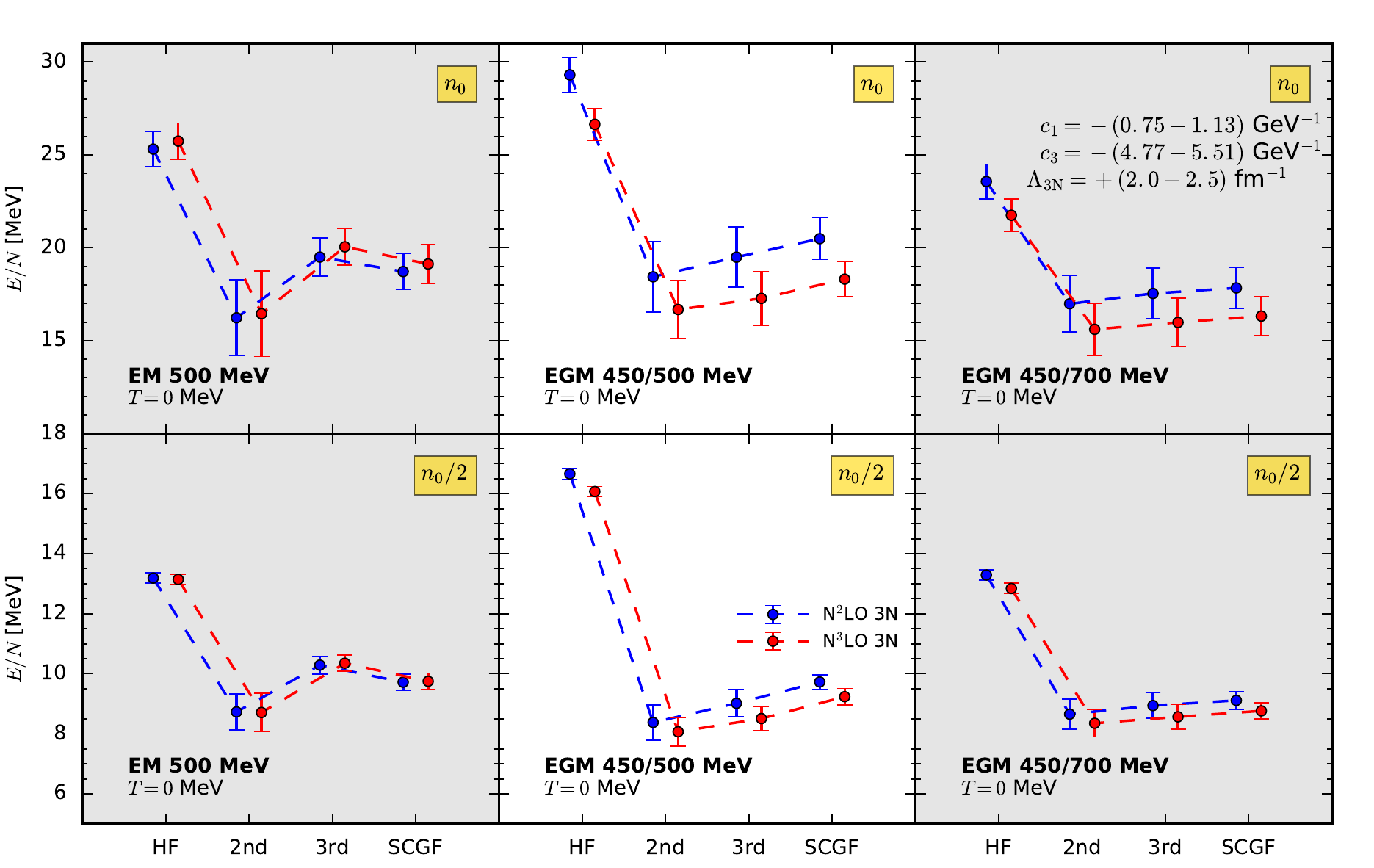}
\caption{\label{fig:MBConv}(Color online)
The energy per particle at different orders of MBPT is shown, up to
Hartree-Fock ($E_{\text{tot}}^{\text{HF}}/N$), second order
($E_{\text{tot}}^{(2)}/N$) and third order ($E_{\text{tot}}^{(3)}/N$),
respectively, in comparison to the energies obtained from the SCGF
method ($E_{\rm SCGF}/N$) at $n_0$ (first row) and $n_0/2$ (second
row), respectively.  The N$^3$LO NN potentials are given in each
panel. Three-body effects are included at N$^2$LO (blue) and at
N$^3$LO (red), respectively. The dashed lines connecting the data
points are in order to guide the eyes. The error bars are due to the
$c_i$ and $\Lambda_\text{3N}$ variations. In this plot, the
third-order calculation does not include the additional many-body
uncertainty (the light-blue band in Fig.~\ref{fig:EOS}).}
\end{figure*}

We show in Fig.~\ref{fig:EOS} the energy per particle as a function of
density in neutron matter at zero temperature. From left to right, the
first row shows the results for the N$^3$LO NN potentials
EM~500~MeV~\cite{Ente03EMN3LO}, EGM~450/500~MeV and
EGM~450/700~MeV~\cite{Epel05EGMN3LO} with leading N$^2$LO 3N
forces. The momentum scales attached to the
potentials correspond to different regulator cutoffs: first, the
cutoff in the Lippmann-Schwinger equation and second, if not
dimensionally regularized, the cutoff in the two-pion-exchange
spectral-function regularization. Analogously, the second row shows the results for the same NN
potentials but including 3N forces up to N$^3$LO.  We consider two
sources of uncertainties: from the chiral Hamiltonian and from
considering only a finite order in MBPT. As stated in
Fig.~\ref{fig:EOS}, the theoretical uncertainties due to the
Hamiltonian are estimated by parameter variation in the 3N forces,
i.e., the cutoff $\Lambda_\text{3N}$ and the low-energy constants
$c_1$ and $c_3$. The $c_i$ values need to be refit at each chiral
order, however, to investigate the net effect of N$^3$LO forces, we
take here solely the $c_i$-range recommended for N$^3$LO
calculations~\cite{Kreb123Nlong}. In addition to the uncertainties in
the Hamiltonian, we estimate the neglected higher-order contributions
in the many-body expansion by varying the single-particle energies at
third order using a free and a Hartree-Fock spectrum. These bands are
colored in dark blue in Fig.~\ref{fig:EOS}. Moreover, following
Ref.~\cite{Krue13N3LOlong} we include the results at second order in
MBPT using a Hartree-Fock spectrum to the uncertainty estimate. This
extension of the pure third-order equation of state is indicated by
light-blue bands. In summary, for a given Hamiltonian we perform in
total three calculations in MBPT: two third-order calculations using
the two single-particle spectra and a second-order calculation using a
Hartree-Fock spectrum. Light- and dark-blue bands together
characterize the total uncertainty estimate of MBPT in each panel. The
actual energy range of MBPT is given in each panel of
Fig.~\ref{fig:EOS} at saturation density $n_0$ (dashed vertical line),
with \mbox{$n_0=0.16$~fm$^{-3}$}.

Let us focus on the results with leading 3N forces, as shown in the
first row of Fig.~\ref{fig:EOS}. The qualitative description does not
change for the calculations with subleading 3N forces (second row in
Fig.~\ref{fig:EOS}).  Whereas the results for the two EGM potentials
are almost independent of the many-body details, the effects of the
variation of spectra and many-body order in MBPT are much more
pronounced for EM~500~MeV: at saturation density the many-body
uncertainties provide contributions of about $\sim -2.5$~MeV for this
Hamiltonian (see light-blue band in Fig.~\ref{fig:EOS}). Including
subleading 3N forces leads basically only to an overall shift of the
bands as shown by the given energy range at saturation density. More
specifically, the net 3N contribution leads to more attraction for the
EGM potentials while the effect on EM~500~MeV is slightly repulsive.

To quantify the many-body convergence in more detail we compare to the
results obtained in the SCGF method which are given by the region
between the red-dashed lines in Fig.~\ref{fig:EOS}. The results in
SCGF are considered to be converged in the many-body expansion (at the
ladder level) and thus include only the uncertainty due to the
Hamiltonian (including variations of the low-energy constants $c_1,
c_3$). We focus again on the different NN potentials rather than on
discussing the effect of subleading 3N forces. Considering the total
uncertainty estimate of MBPT we find for the potentials EM~500~MeV and
EGM~450/700~MeV completely overlapping bands and similar trends in
density. In the case of EM~500~MeV the extended uncertainty
(light-blue band) is however needed to obtain more attraction and
consequently fully overlapping bands, whereas for EGM~450/700~MeV the
pure third-order energy is already in remarkable agreement. In
addition to the above discussion on the size of the light-blue bands
this suggests that contributions beyond third-order are small for
EGM~450/700~MeV and become significant for EM~500~MeV.

For EGM~450/500~MeV we observe a slightly different density dependence
between the MBPT and the SCGF curves, leading to an almost total
overlap at saturation density but less agreement in the region around
$n \sim 0.1$~fm$^{-3}$. Here, the equation of state in SCGF is
slightly more repulsive. We recall that the SCGF results are
extrapolated down to zero temperature from calculations performed at
$T=2$~MeV for $n \leqslant 0.05$~fm$^{-3}$ and at $T=5$~MeV for
densities above.  We have tested whether this discrepancy is related
to the extrapolation to zero-temperature lowering the temperature down
to $T=3,4$~MeV in densities between 0.05 and 0.10~fm$^{-3}$, and have
found no dependency on the extrapolation.

Combining the discussions on the size of the additional many-body
uncertainty and the comparison of MBPT vs. SCGF we conclude from
Fig.~\ref{fig:EOS} that the perturbativeness improves from EM~500~MeV
to EGM~450/500~MeV to EGM~450/700~MeV. It is remarkable that a
third-order MBPT calculation compares so well with the nonperturbative
case for these chiral NN potentials. We study the many-body
convergence as well as the effect of subleading 3N forces in more
details in the next section.

\subsection{Many-body convergence}

In Fig.~\ref{fig:MBConv} we address again the many-body convergence
and show order-by-order in MBPT the total energy per neutron at $n_0$
(first row) and $n_0/2$ (second row), analogously to
Fig.~\ref{fig:EOS}. More specifically, we show the total energy in
Hartree-Fock approximation $E_\text{tot}^{\text{(HF)}}/N$ (``HF"),
second order (``2nd") and third order (``3rd"), $E_\text{tot}^{(2)}/N$
and $E_\text{tot}^{(3)}/N$ respectively, in comparison to the results
obtained in the SCGF method, $E_{\rm SCGF}/N$ (``SCGF"). The
uncertainties are obtained as in Fig.~\ref{fig:EOS} through variations
of the 3N parameters and the single-particle energies. However, to
study the many-body convergence the third-order bands do not include
here the additional many-body uncertainty (the light-blue bands of
Fig.~\ref{fig:EOS}). The blue (red) data points correspond to N$^2$LO
(N$^3$LO) 3N forces.

For all six panels in Fig.~\ref{fig:MBConv} we observe similar overall
patterns: comparing order-by-order to the SCGF method we observe that
the second order adds always too much attraction which then is
compensated by the third-order repulsion.  However, the specific
behavior is different for EM~500~MeV and the two EGM potentials. In
the case of EM~500~MeV the large third order overcompensates the
second-order repulsion. In contrast, the third-order contribution is
much smaller and less repulsive for the EGM potentials as can be seen
in Fig.~\ref{fig:MBConv} (second and third column).  In particular,
this is pronounced in the calculations based on EGM~450/700~MeV, which
agree remarkably well with the SCGF result.

As already discussed in the description of Fig.~\ref{fig:EOS},
including N$^3$LO 3N forces has only a small repulsive effect on the
energies based on EM~500~MeV, whereas the effect on the EGM potentials
is larger but attractive.  This behavior can be traced back to NN-3N
mixing terms that enter the calculation when including 3N forces
beyond the HF level. We also note that the values of the low-energy
constants $C_S$ and $C_T$, which enter N$^3$LO 3N contributions,
differ for all three potentials. However, the many-body convergence is
not altered by including contributions from subleading 3N
interactions.

\subsection{Comparison to previous calculations at N$^3$LO}

\begin{figure*}[t]
\includegraphics[page=1,scale=0.95,clip]{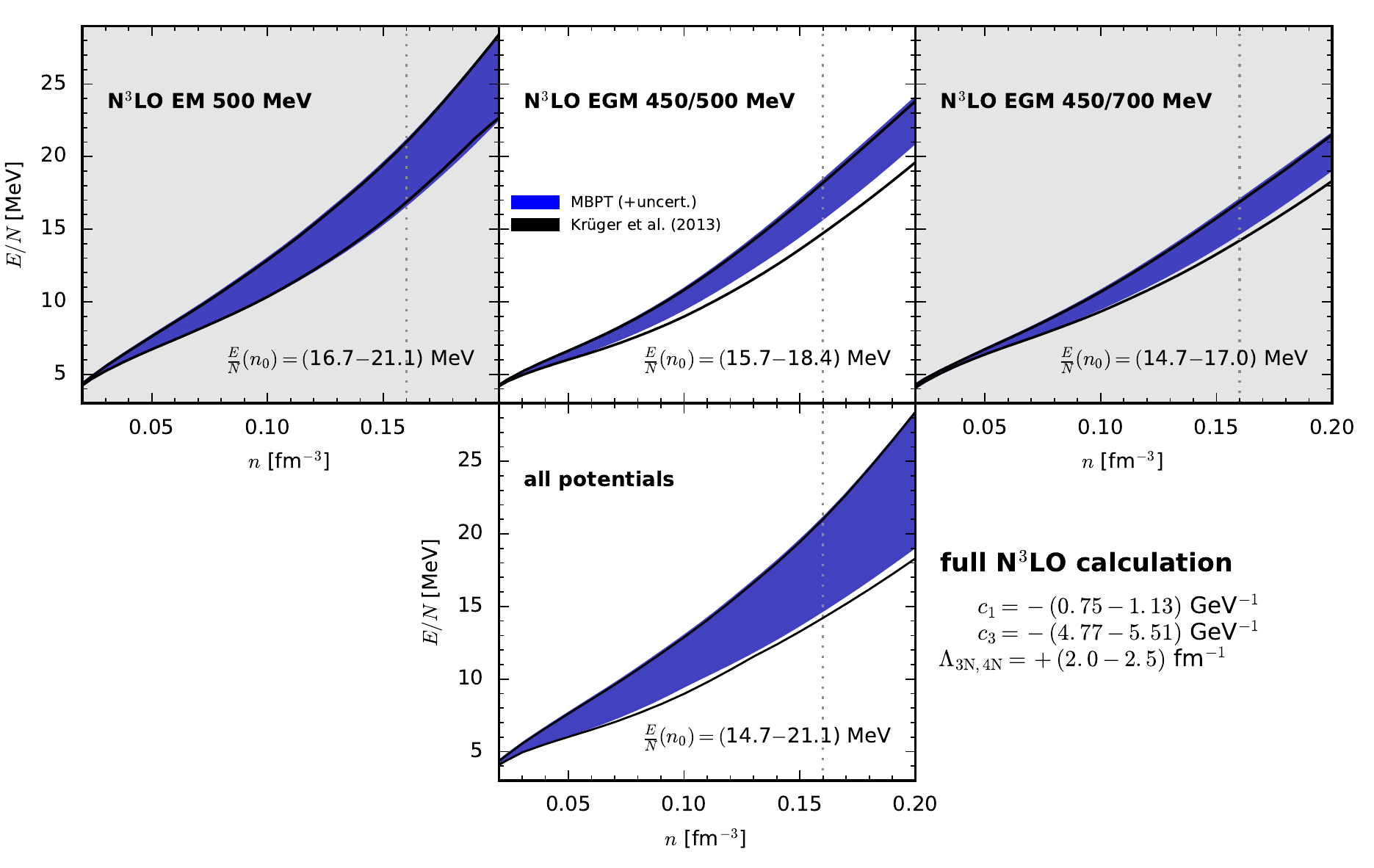}
\caption{\label{fig:fullN3LO}(Color online)
The energy per particle of neutron matter at N$^3$LO for the three
different NN potentials (this work: blue bands) in comparison to
Ref.~\cite{Krue13N3LOlong} (Kr\"uger et al. (2013): black lines). The
second row combines the results of the first row. In each panel, we
give the energy range at saturation density obtained within the
improved calculations presented in this work. See text for details.}
\end{figure*}

The authors of Refs.~\cite{Tews13N3LO,Krue13N3LOlong} performed the
first consistent calculations at N$^3$LO including NN, 3N and 4N
forces in MBPT. In the cited works N$^3$LO NN and N$^2$LO 3N forces
have been considered up to third order in MBPT in terms of effective
NN potentials~\cite{Hebe10nmatt}, whereas subleading 3N interactions
could only be included in the Hartree-Fock approximation since no 3NF
partial-wave matrix elements were available at that time. Thanks to
the advances discussed in this paper we are now in the position to
revisit and systematically improve these calculations.

In Fig.~\ref{fig:fullN3LO} we show our improved results for the energy
of neutron matter (blue bands) for the three Hamiltonians EM~500~MeV,
EGM~450/500~MeV and EGM~450/700~MeV (first row) and the total band
merged from the previous panels (second row). The uncertainty bands
cover again variations of the 3N parameters (as given in the figure),
the single-particle spectrum and the additional many-body uncertainty
(see also discussion of Fig.~\ref{fig:EOS}).  We furthermore include
the 4N Hartree-Fock results, as given in Ref.~\cite{Krue13N3LOlong},
and vary the 4N cutoff analogously to the 3N forces. In addition, we
show the results of Ref.~\cite{Krue13N3LOlong}\footnote{For completeness, 
we have corrected a small error in the routines of
Ref.~\cite{Krue13N3LOlong} for the computation of the second- and
third- order contribution of the N$^3$LO NN plus N$^2$LO 3N forces as
well as the N$^3$LO 3N Hartree-Fock energy corresponding to the ring
topology. Moreover, we are using the typo-corrected values for
$\bar{\beta}_{8,9}$ (see Ref.~\cite{Epel15improved} for details).}
depicted by the black solid lines.  For a better view we do not fill
this region. We give in each panel the energy range at saturation
density obtained within the improved calculations presented in this
work.

We observe that the effect of adding the N$^3$LO 3N contributions
beyond Hartree-Fock varies significantly between the EM~500~MeV and
the two EGM potentials. For EM~500~MeV these contributions leave the
uncertainty band almost unaffected. For the two EGM potentials the
upper uncertainty limits remains the same while the lower increase by
$\sim1$~MeV ($\sim0.2$~MeV) for EGM~450/500~MeV (EGM~450/700~MeV),
hence decreasing the width of the uncertainty band. These findings are
consistent with the observations in Ref.~\cite{Krue13N3LOlong}, which
stated that the N$^3$LO 3N Hartree-Fock energy is smaller for
EM~500~MeV while it is much larger for the two EGM potentials (see
Fig.~6 of Ref.~\cite{Krue13N3LOlong}). We emphasize, however, that NN
and effective NN forces get mixed at second-order and beyond, and
therefore the net effect of these subleading 3N contributions cannot
be easily disentangled in the many-body calculation. Combining all
bands we find a total uncertainty of $\frac{E}{N} (n_0) = (14.7 -
21.1)~\text{MeV}$ in neutron matter at saturation density. Compared to
the corrected total band of Ref.~\cite{Krue13N3LOlong} $\frac{E}{N}
(n_0) = (14.3 - 21.1)~\text{MeV}$, we obtain a slight reduction of the
lower limit of the uncertainty band. As suggested in
Refs.~\cite{Tews13N3LO,Krue13N3LOlong}, these effects are indeed
rather small. However, we expect the effects to be much more important
as soon as the proton fraction is finite (see also discussion of
symmetric nuclear matter in Ref.~\cite{Krue13N3LOlong}).

\subsection{\label{subsec:NO_finT}Normal-ordering at finite temperatures}

\begin{figure}[t]
\includegraphics[page=1,scale=1.1,clip]{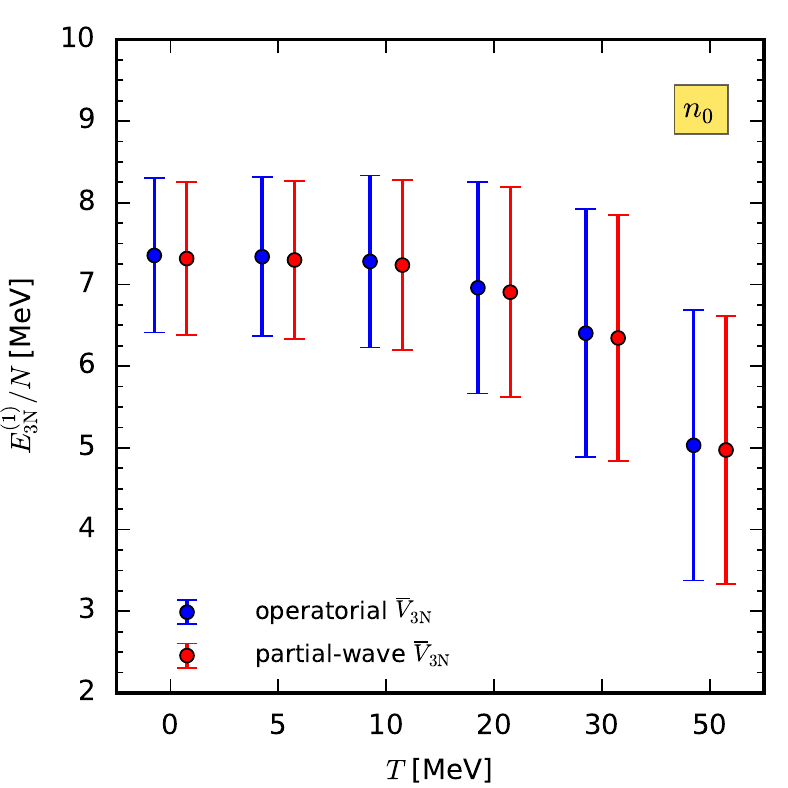}
\caption{\label{fig:finiteT_HF}(Color online)
Comparison of the leading 3N Hartree-Fock energies at saturation
density for several temperatures obtained using the effective NN
potential in terms of 3N operators (blue) and the partial-wave
approach (red). We include 3N matrix elements up to $\mathcal{J}
\leqslant 9/2$ and $J \leqslant 6$. For the uncertainty estimate we
use the same parameter variation in the 3N forces as in
Fig.~\ref{fig:EOS}.}
\end{figure}

We have extended the recently-developed framework for computing
effective NN potentials in a partial-wave basis~\cite{Dris16asym} to
finite temperatures.  Besides being a necessary step in order to
include these matrix elements in the SCGF method (due to the
extrapolation from finite temperatures), this is also a crucial step
for future MBPT calculations of nuclear matter at finite
temperatures. In Fig.~\ref{fig:finiteT_HF} we show the resulting
N$^2$LO 3N Hartree-Fock energies $E^{(1)}/N(n_0,T)$ at six different
temperatures in the range of $T=(0-50)$~MeV. We benchmark our new
values (red) against previous results (blue) obtained via an
operatorial approach~\cite{Carb14SCGFdd}. The uncertainty bands are
obtained through 3N parameter variation analogously to
Figs.~\ref{fig:EOS} and~\ref{fig:MBConv}.  The single-particle
spectrum does not contribute to the uncertainties since the
Fermi-Dirac distribution in Eq.~\eqref{eq:Veff_formal} is computed
using a free spectrum. A similar benchmark at N$^3$LO is not possible
since no matrix elements are currently available based on the
operatorial evaluation of 3N forces at N$^3$LO. We note that the 3N
interaction energy decreases with temperature as shown in
Fig.~\ref{fig:finiteT_HF}. Including also kinetic energy contributions
would lead to a total increase in energy with increasing
temperature. From Fig.~\ref{fig:finiteT_HF} we can conclude that the
two different methods for the normal-ordering agree very well at zero
and finite temperature up to $T=50$ MeV.

\begin{figure}[t]
\includegraphics[page=1,scale=1.1,clip]{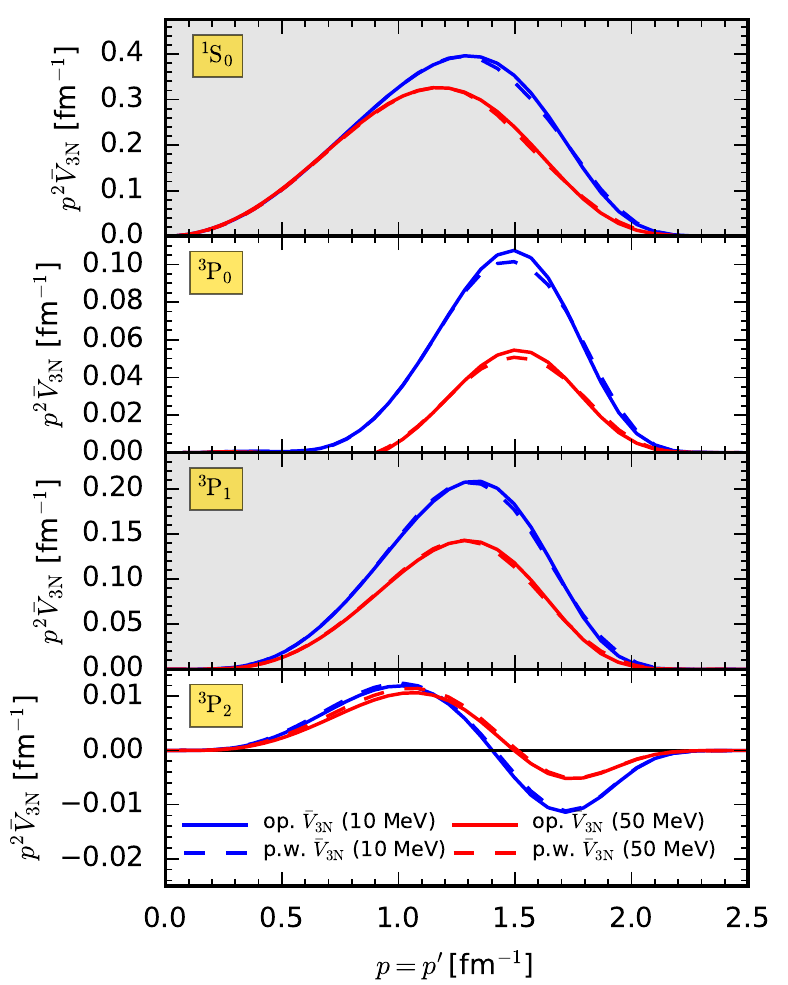}
\caption{\label{fig:Veff_T}(Color online)
Momentum-space diagonal matrix elements of the density-dependent 
effective NN potentials at N$^2$LO for a selection of four
partial-wave channels and two temperatures.}
\end{figure}

In addition to the 3N Hartree-Fock energies, we also benchmark the
underlying interaction matrix elements of the effective potential
$\overline{V}_\text{3N}^\text{as}$. The results for a selection of
four partial-wave channels and two temperatures are shown in
Fig.~\ref{fig:Veff_T}.  These matrix elements contribute to the energy
presented in Fig.~\ref{fig:finiteT_HF}. The ones obtained in the partial-wave 
(operatorial) approach are plotted as dashed (solid) lines. 
We select a representative set of channels, ${}^1\text{S}_0$,
${}^3\text{P}_0$, ${}^3\text{P}_1$, and ${}^3\text{P}_2$, and
temperatures $T=10,50$~MeV.  We have also compared higher partial
waves up to $J=6$ and momentum off-diagonal matrix elements for
$\Lambda_\text{3N}=(2.0-2.5)$~fm${}^{-1}$. As in
Ref.~\cite{Dris16asym}, we find indications of an incomplete
partial-wave convergence only for partial-waves channels with $J>4$,
We also checked that the agreement can be systematically improved by
increasing the 3N model space, i.e. by including channels with
$\mathcal J =11/2$ and $13/2$. We found that contributions from these
higher partial-wave channels provide $\lesssim 50$~keV to the energy
of neutron matter per particle at saturation density. Overall, we find
excellent agreement of the two methods at the level of matrix elements
and at finite-temperatures. This shows that the computed matrix
elements of the effective interactions at finite temperature at
N$^2$LO and N$^3$LO are correct and numerically stable and are hence
suitable for future calculations of nuclear matter for astrophysical
applications~\cite{Hebe13ApJ}.

\section{\label{sec:outlook}Summary and Outlook}

In this work we have calculated the zero-temperature equation of state
of neutron matter in the framework of MBPT and SCGF based on chiral NN
and 3N forces up to N$^3$LO. In addition, we included contributions
from 4N interactions at N$^3$LO in the Hartree-Fock approximation. For
the inclusion of 3N interactions we have utilized our generalized
normal-ordering framework first presented in
Ref.~\cite{Dris16asym}. We demonstrated that this framework is able to
treat general 3N interactions that are provided in a partial-wave
representation and can be extended to finite temperatures.

We have systematically improved previous calculations of neutron
matter in MBPT at N$^3$LO~\citep{Tews13N3LO,Krue13N3LOlong} by
including subleading 3N contributions beyond the Hartree-Fock
approximation.  Specifically, we have obtained the neutron-matter
energy based on three different NN plus 3N interactions derived within
chiral EFT, comparing calculations including only leading to
up-to-subleading 3N forces.  For the N$^3$LO NN potentials
EGM~450/500~MeV and EGM~450/700~MeV we found additional attractive
subleading 3N contributions of about $\sim$2~MeV for the energy per
particle at saturation density, while for the EM~500~MeV potential
these contributions are smaller in size and repulsive, of the order of
$\sim$500~keV. In order to assess the many-body convergence we have
benchmarked our MBPT results for three commonly-used N$^3$LO NN
potentials plus leading and also subleading 3N forces against results
obtained within the SCGF framework. Since the current implementation
of SCGF does not account for Cooper pairing, the zero-temperature
limit was obtained by extrapolation. We found a systematic convergence
of the MBPT results to the SCGF results at third order in MBPT,
whereas the detailed convergence pattern depends on details of the NN
and 3N interactions.

Finally, we have successfully benchmarked results for the effective NN
potential at finite temperature. At order N$^2$LO in the chiral
expansion we obtain excellent agreement between results obtained using
our novel normal- ordering framework and previous results for 3N
Hartree-Fock energy contributions as well as on the level of
partial-wave matrix elements. These benchmarks demonstrate that we are
now in the position to perform calculations of general
isospin-asymmetric matter including all NN and 3N contributions up to
N$^3$LO at zero and finite temperatures. Since all 3N topologies
contribute for these systems, reliable fits of the 3N low-energy
constants $c_D$ and $c_E$ are required. This is currently work in
progress.  The availability of different sets of Hamiltonians using
different regulator choices (see also
Refs.~\cite{Tews15QMCPNM,Dyhd16Regs}) and different fitting strategies
(see, e.g., Refs.~\cite{Ekst15sat,Carl16sim}) will make it possible to
probe systematically the order-by-order convergence in the chiral
expansion.  In turn, this will advance our understanding of the dense
matter equation of state.

\begin{acknowledgments}
We thank T. Kr{\"u}ger, A. Rios, A. Polls and I. Tews for useful discussions.
This work was supported by the ERC Grant No.~307986 STRONGINT, the Deutsche
Forschungsgemeinschaft through Grant SFB 1245. A.C. acknowledges support by
the Alexander von Humboldt Foundation through a Humboldt Research Fellowship
for Postdoctoral Researchers.
\end{acknowledgments}

\bibliographystyle{apsrev4-1}
\bibliography{strongint}

\end{document}